\documentclass[pra,twocolumn,
 superscriptaddress,amsmath,amssymb,nofootinbib]{revtex4-2}
\usepackage{amsfonts}
\usepackage{amsmath}
\usepackage{bbm}
\usepackage{amssymb}
\usepackage{graphicx}%
\usepackage{footmisc}
\usepackage{bm}
\usepackage{braket}
\usepackage{hyperref}
\usepackage{cleveref}
\usepackage{longtable,booktabs}
\usepackage{tikz}
\usepackage[all,arc,knot,matrix]{xy}
\crefname{equation}{Eq.}{Eqs.}

\newcommand{\abs}[1]{\lvert#1\rvert}

\newcommand{\Tr}{\text{Tr}}

\begin{document}
\title{Leggett-Garg inequalities for multitime processes}

\author{Zhiqiang Huang }
\email{hzq@wipm.ac.cn}
\affiliation{State Key Laboratory of Magnetic Resonance and Atomic and Molecular Physics, Innovation Academy for Precision Measurement Science and Technology, Chinese Academy of Sciences, Wuhan 430071, China}
\author{Xiao-Kan Guo }
\email{kankuohsiao@whu.edu.cn}
\affiliation{Department of Applied Mathematics, Yancheng Institute of Technology, Jiangsu 224051, China}

\date{\today}

\begin{abstract}
 We study some aspects of the Leggett-Garg inequalities by using the operator-state formalism for multitime processes. The process tensor in its Choi-state form, which we call process state, is employed to investigate the Leggett-Garg  inequalities and their violations. We find the sufficient conditions on process states for the Leggett-Garg inequalities to hold, based on which we find a new way of characterizing the  influences on the violation of Leggett-Garg inequalities through the  structure of process states.
\end{abstract}


\maketitle

\section{Introduction}\label{INTRO}
The Bell inequalities or Clauser-Horne-Shimony-Holt (CHSH) inequalities \cite{Bell,CHSH} serve as fundamental tests for local realism in physical theories. Quantum theory's violation of these inequalities demonstrates its departure from local hidden-variable theories, revealing inherent nonlocal quantum correlations. Fine's theorem \cite{Fine} establishes that the CHSH inequalities represent both necessary and sufficient conditions for local realism, making their violation an unambiguous indicator of nonclassical correlations.

The temporal counterpart to CHSH inequalities emerges in Leggett-Garg (LG) inequalities \cite{LG}, which probe macroscopic realism (macrorealism) rather than local realism. While CHSH inequalities require device independence, LG inequalities relax this constraint. Macrorealism postulates that physical systems possess definite properties independent of measurement, akin to classical macroscopic objects. LG inequalities specifically examine temporal correlations in sequential measurements on single systems, where violations indicate non-macrorealistic behavior. Unlike the CHSH case, no analog of Fine's theorem exists for LG inequalities \cite{CK16}, as they only constitute necessary conditions for macrorealism. Subsequent research has sought complete characterizations through no-signaling in time conditions \cite{NSIT1,NSIT2} and augmented inequality sets \cite{Hal1,Hal2}. Recent experimental advances, including loophole-free interferometric tests with heralded single photons \cite{JSHS22}, have conclusively demonstrated LG violations while addressing historical challenges like clumsiness and detection loopholes.

Open quantum systems provide a natural framework for LG inequality studies due to their inherent time evolution and measurement interactions. However, environmental influences introduce additional complexities: both Markovian \cite{Ema13,FL17,CDMSS18,GVD21} and non-Markovian dynamics \cite{CA14,DMM18,Ban19,NBS20} significantly affect LG violation patterns. Environmental conditions (equilibrium vs. non-equilibrium) further modulate these effects \cite{CRQ13,LLD15,ZWW20}, while hybrid influences combining system dynamics and environmental interactions create novel phenomena \cite{Ban18,Hidden}. These factors may invalidate standard LG inequality assumptions, complicating the interpretation of violations as direct macrorealism tests. Nevertheless, understanding such influences remains crucial for experimental implementations.

Given this complex landscape of modifying factors, recent work \cite{Plenio19} establishes a direct connection between LG inequality violations and quantum coherence for Markovian processes, preserving their role as nonclassicality witnesses. Extensions to non-Markovian regimes \cite{Plenio20} employ process tensors – powerful operational tools for characterizing temporally extended quantum processes \cite{PT,PTR2}. Building on these developments, our work employs the operator-state formalism to investigate LG inequalities through process tensor analysis.

The operator-state formalism, partially developed in \cite{PCASA19,Huang22}, treats quantum operators as states in an enlarged Hilbert space, enabling Choi-state representations of multitime processes (hereafter termed ``process states"). This framework facilitates derivation of sufficient conditions for LG inequality satisfaction through ``quantum-classical" process state constraints. Violations emerge from deviations from these constrained forms, explicitly incorporating state disturbance and non-Markovian effects.

This paper is organized as follows. In section \ref{S2}, we first introduce the general framework of operator states and rewrite the Choi-state form of the process tensors, or process states, in this framework.
Then we rewrite the LG inequalities for two-time measurements using the two-time probabilities computed from the process states.
 In section \ref{S3}, we derive a set of sufficient conditions on the process states for the LG inequalities to hold. When these conditions do not hold, we obtain modified LG inequalities.
  In section \ref{S4}, the influences on the open system are considered by comparing different process states/tensors with those satisfying the the sufficient conditions.
 Section \ref{S6} concludes.

\section{ LG inequalities using process states}\label{S2}

\subsection{Operator states and process states}
Let us introduce the operator state  $|O)$ for an operator $O$ on the Hilbert space $\mathcal{H}_S$ of quantum states of a system $S$, which belongs to a new  Hilbert space $\mathsf{H}$ with an inner product defined by 
\begin{equation}
    (O_1|O_2)=\text{Tr}(O_1^\dag O_2).
\end{equation}
  An orthonormal basis for $\mathsf{H}$ is $|\Pi_{ij})$ where $\Pi_{ij}=\ket{i}\bra{j}$, as one can check that 
  \begin{equation}
  (\Pi_{kl}|\Pi_{ij})=\delta_{ik}\delta_{jl}.
  \end{equation}
  The completeness of this basis $\{|\Pi_{ij})\}$ is
    \begin{equation}\label{3}
{\bf1}_\mathsf{H}=\sum_{ij}|\Pi_{ij})(\Pi_{ij}|.
  \end{equation}
  
For arbitrary operators $O_i$, we have the conjugate relation
\begin{equation}\label{c4}
    [(O_1|O_2)]^\dagger=\text{Tr}(O_1^\dag O_2)^\dagger=\text{Tr}(O_2^\dag O_1)=(O_2|O_1).
\end{equation}
Therefore, we can define the Hermitian conjugate of the operator state as $|O)^\dag:=(O|$. Using the permutation property of trace, we have
\begin{equation}\label{perpt}
    [(O_1|O_2)]^\dagger=\text{Tr}(O_1 O_2^\dag )=(O_1^\dagger|O_2^\dagger).
\end{equation}
If the Hilbert spaces of two operator states are inconsistent, but there is a common subspace under the tensor decomposition, such as $\mathcal{H}_1=\mathcal{H}_C\otimes \mathcal{H}_A$ and $\mathcal{H}_2=\mathcal{H}_C\otimes \mathcal{H}_B$, then the inner product is defined as follows
\begin{equation}
    (O_1|O_2)=\text{Tr}_{\mathcal{H}_C}(O_1^\dag O_2),
\end{equation}
which is still an operator state belonging to $ \mathcal{L}(\mathcal{H}_A\otimes \mathcal{H}_B)$. It should be noted that when the Hilbert spaces of the two operator states are inconsistent, \cref{c4} still holds, but \cref{perpt} no longer holds.
By applying the completeness relation \eqref{3}, we can gradually expand:
\begin{align}
   ({O}^S|=&\sum_{ij}({O}^S|\Pi_{ij}^S)( \Pi_{ij}^S|=\nonumber\\
   =&\sum_{ij}[(\Pi_{ij}^A |{O}^A)]^\dagger( \Pi_{ij}^S|=\notag \\
   =&\sum_{ij}(\Pi_{ji}^A \otimes \Pi_{ij}^S|{{O}^A}^\dagger)=\notag \\
   =&\sum_{ij}(\Pi_{ij}^A \otimes \Pi_{ij}^S|{{O}^A}^*)\equiv (\Phi^{AS}|{{O}^A}^*)\label{c5}
\end{align}
where in the second line we have changed the index from $S$ to $A$ without loss of generality and used the conjugate relation (\ref{c4}), in the third line we have used the inner product property \eqref{perpt} and in the last line the matrix indices of $\Pi_{ji}$ has been changed to $\Pi_{ij}$ so that the Hermitian conjugate $\dag$ has been changed to the complex conjugate $*$. Here, $\Phi^{AS}=\ket{\phi}\bra{\phi}$ is the density operator of maximally entangled state $\ket{\phi}=\sum_i \ket{i}_A\otimes\ket{i}_S $ in $\mathcal{H}_A\otimes\mathcal{H}_S$, as $A$ and $S$ label the identical system but with (artificially) different indices. The  $\Phi^{AS}$ is a  Hermitian operator satisfying ${\Phi^{AS}}^\dag=\Phi^{AS}$.  The normalization of $\Phi$ and $\phi$ is hidden for simplicity. According to the Hermitian conjugate of the operator state and \cref{c5}, we obtain $ |{O}^S)=({{O}^A}^*|\Phi^{AS})$.
In effect, we have doubled the system Hilbert space $\mathcal{H}_S$ to $ \mathcal{H}_A\otimes\mathcal{H}_S$, and then projected out the $A$-part. By introducing the tensor product structure of the auxiliary space $\mathcal{H}_A$ and the system space $\mathcal{H}_S$, we can naturally map the operator $O^S$ on $\mathcal{H}_S$ to the $\mathcal{H}_A$ space through Hermitian conjugation $\dag$. Specifically, for any operator $O\in \mathcal{L}(\mathcal{H}_S)$, the corresponding auxiliary space operator is defined as ${{O}^A}^* = (O^{\dag})^T \in \mathcal{L}(\mathcal{H}_A)$ where $T$ represents the matrix transpose. This mapping preserves the algebraic structure of the operator and satisfies
\begin{align}
  \bra{i}  O^S\ket{j}=  (\Pi_{ij}^S|O^S)=({{O}^A}^*\otimes\Pi_{ij}^S|\Phi^{AS})\notag\\
  =({{O}^A}^*|\Pi_{ij}^A)=\Tr[({O}^A)^T\Pi_{ij}^A]= (\Pi_{ij}^A|O^A)
\end{align}
where we have used $ |{O}^S)=({{O}^A}^*|\Phi^{AS})$ in the second equality. 

 With the help of  \eqref{c5}, we can consider the action of a quantum operation $\mathcal{N}^S$ on the state $\rho^S$ of the system $S$:
\begin{equation}\label{c6}
    |\mathcal{N}^S(\rho^S))= \mathcal{N}^S|\rho^S)=({\rho^A}^*|\mathcal{N}^S|\Phi^{AS})\equiv ({\rho^A}^*|\Psi^{AS}_\mathcal{N})
\end{equation}
where  $\mathcal{N}^S$ can be pulled into the round bracket because $({\rho^A}^*|$ acts only on the $A$ which is treated differently than $S$. The \cref{c6} can be expressed diagrammatically as 
\begin{equation}
    \includegraphics[width=1\linewidth]{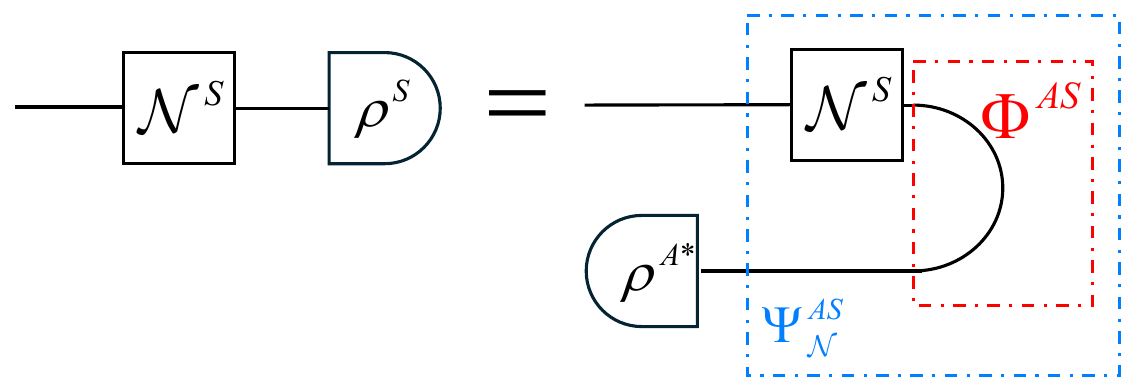} 
\end{equation}    
According to the Choi-Jami\l kowski  isomorphism theorem, any quantum operation $\mathcal{N}^S: \mathcal{L}(\mathcal{H}_S) \to \mathcal{L}(\mathcal{H}_S')$ can be expressed as a state in the dual space. The specific construction goes as follows: Take the auxiliary space of $\mathcal{H}_A \cong \mathcal{H}_S$, use the maximally  entangled state $\Phi^{AS}$, then the Choi state corresponding to $\mathcal{N}^S$ is $|\Psi^{AS}_\mathcal{N})\equiv I_A \otimes \mathcal{N}^S|\Phi^{AS})$.
This is exactly the result shown in equation \eqref{c6}. By applying the quantum operation $\mathcal{N}^S$ to the $S$ part of $\Phi^{AS}$, we obtain the Choi state $\Psi^{AS}_\mathcal{N}$ that represents all the information of $\mathcal{N}^S$. This representation transforms the concatenation of quantum operations into a contraction operation of the Choi state.

Based on the definition of a maximally entangled state and the inner product of operator states, one can readily prove that
\begin{equation}\label{rsinc}
    (\Phi^{S_1A'} |\Phi^{A'S})=\sum_{ij}|\Pi_{ij}^{S_1})( \Pi_{ij}^S|.
\end{equation}
This expression closely resembles the completeness relation, albeit with different operator state indices, and serves to reassign the indices of the operator states.
Following \cref{rsinc}, we can also consider the successive actions of two quantum operations $\mathcal{N^S}$  and $\mathcal{M^S}$:
\begin{align}
   \mathcal{N}^S \circ\mathcal{M}^S|\rho^S)=\mathcal{N}^S \circ  (\Phi^{S_1A'} |\Phi^{A'S})\circ\mathcal{M}^{S_1}|\rho^{S_1})\nonumber\\
    =   (\Phi^{S_1A'} |\mathcal{M}^{S_1}|\mathcal{N}^S(\Phi^{A'S}))|\rho^{S_1})\equiv({\Psi_\mathcal{M}^{S_1A'}}|\Psi^{A'S}_\mathcal{N})|\rho^{S_1}).\label{7}
\end{align}
The strangely looking indices of \eqref{7} are a feature of the operator-state formalism for multitime processes, cf. \cite{Huang22}. 
 In fact, \cref{7}  can be expressed  diagrammatically as
\begin{equation}\label{c8}
    \includegraphics[width=1\linewidth]{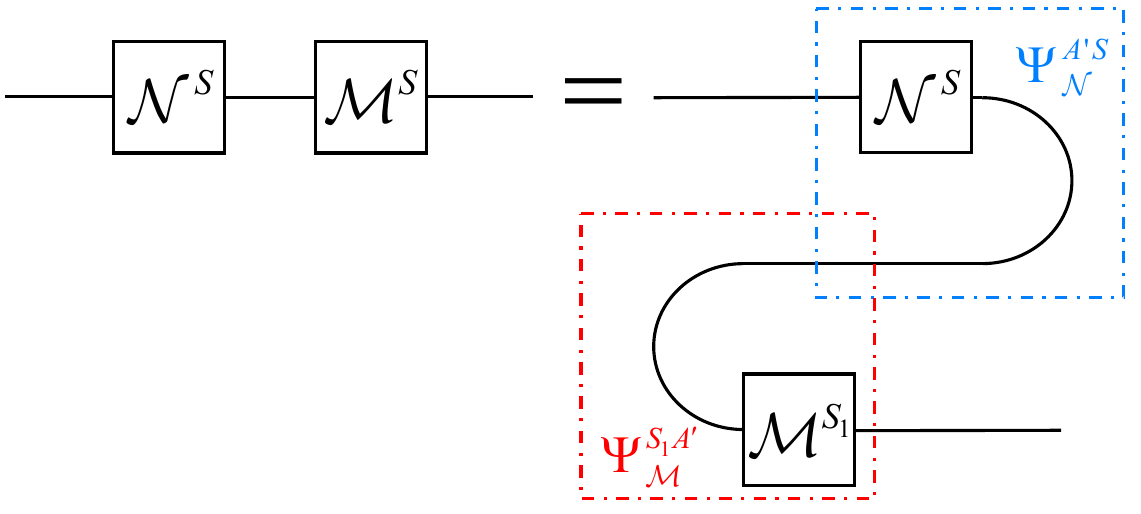} 
\end{equation}    
So, the indices in the right hand side of \eqref{7} are actually telling us the same story as the left hand side of \eqref{7} with the help of two maximally entangled states (or Choi states).

For more actions of quantum operations, the expression will become more complicated. However, for computing the probabilities of multitime processes, we only need to use \eqref{7}. Indeed, we can compute the trace of the state obtained from three quantum operations $\mathcal{M}_i,i=1,2,3$ intersected with two unitary time evolutions $\mathcal{U}^{SE}$ of the initial state $\rho_0^{SE}=\rho_0^S\otimes\rho_0^E$ 
(shown here under the assumption of no initial system-environment correlations; to include such correlations, replace the factorized state with a generic $\rho_0^{SE}$ in the process states) as
\begin{widetext}
\begin{align}
    \mathcal{P}_{3:1}= &(I^S\otimes I^E|\mathcal{M}_3^{S}\circ \mathcal{U}_2^{SE} \circ\mathcal{M}_2^{S}\circ \mathcal{U}_{1}^{SE}\circ\mathcal{M}_1^S|\rho_0^S\otimes \rho_0^E)=\nonumber\\
    =&(I^S\otimes I^E|\mathcal{M}_3^{S}\circ \mathcal{U}_2^{SE} \circ\mathcal{M}_2^{S}
    \left[(\Phi^{S_1A'}|\mathcal{M}_1^{S_1} |\rho_0^{S_1}\otimes \mathcal{U}_1^{SE} [\Phi^{A'S}\otimes\rho_0^E])\right] =    \notag \\
    =&(I^S\otimes I^E|\mathcal{M}_3^{S}
    (\Phi^{S_2A''}\otimes\Phi^{S_1A'} |\mathcal{M}_2^{S_2}\otimes\mathcal{M}_1^{S_1} |\rho_0^{S_1}\otimes  \mathcal{U}_2^{SE}\circ \mathcal{U}_1^{S_2E} [\Phi^{A''S}\otimes\Phi^{A'S_2}\otimes\rho_0^E])=\notag \\
   = &(I^{S_3}\otimes \Phi^{S_2A''}\otimes \Phi^{S_1A'}|\mathcal{M}_3^{S_3}\otimes\mathcal{M}_2^{S_2}\otimes\mathcal{M}_1^{S_1}(I^E|  \mathcal{U}_2^{S_3E}\circ \mathcal{U}_1^{S_2E}|\rho_0^{S_1}\otimes [\Phi^{A''S_3}\otimes\Phi^{A'S_2}\otimes\rho_0^E]),\label{9}
\end{align}
which  can be expressed  diagrammatically as
\begin{equation}
    \includegraphics[width=0.8\linewidth]{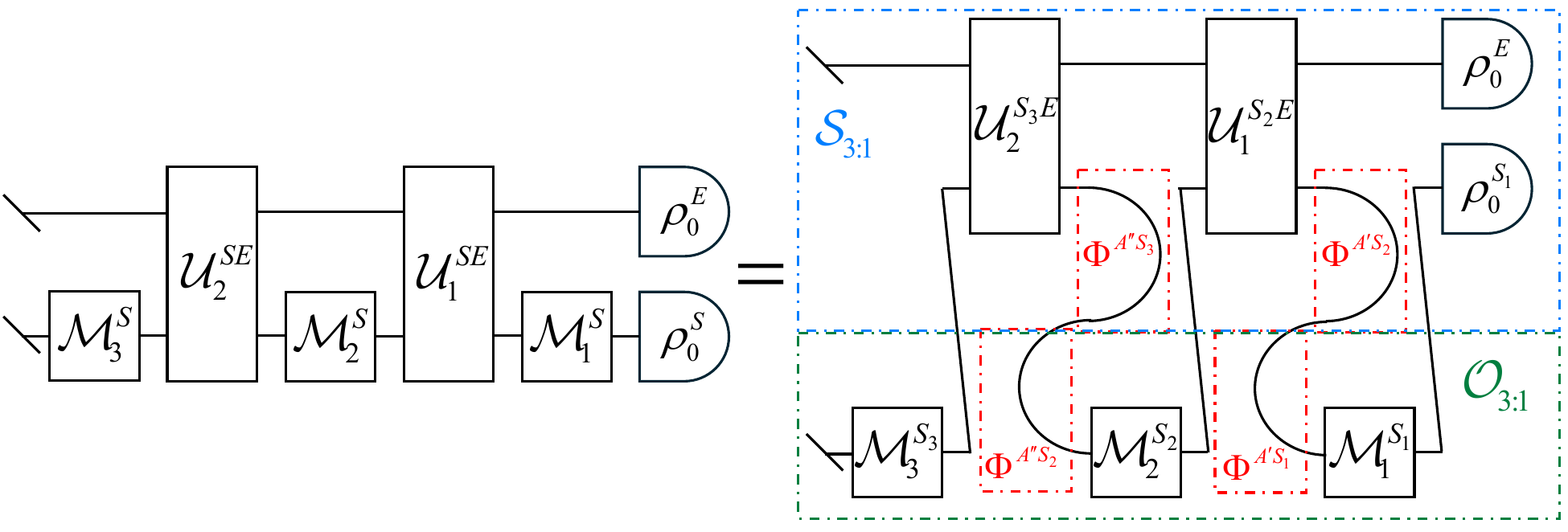} 
\end{equation}    
The labels $S_i,i=1,2,...,n$ (which are ordered as $S_{1}\rightarrow S_{2}\rightarrow\ldots\rightarrow S_{n}$) labels the input state for each operation. 
We observe that \eqref{9} contains two parts: one part is the evolution of the state (on the right) and the other is the quantum operations (on the left). Diagrammatically, this type of algebraic manipulation means connecting $(n-1)$ segments of the curved lines of \eqref{c8}.
By repeating this  manipulation up to the $n$-th operation, we can obtain the corresponding $n$-point probability.
In the general $n$-point case, we obtain the {process state} as
\begin{equation}\label{10}
    |\mathcal{S}_{n:1}):= (  I^E|\mathcal{U}^{S_{n} E}_{n-1}\circ \ldots\circ \mathcal{U}^{S_2 E}_1|\rho_0^{S_1}\otimes [\bigotimes_{j=2}^{n}  \Phi^{A^{(j-1)}S_{j}}]\otimes\rho_0^E). 
\end{equation}\end{widetext}
We also define the $n$-point operation as
\begin{equation}\label{11}
   (\mathcal{O}_{n:1}^{(x_n:x_1)}|:= (I^{S_n}\otimes[\bigotimes_{i=1}^{n-1}  \Phi^{S_{i}A^{(i)}}]|\mathcal{M}_n^{(x_n)}\otimes  \ldots \otimes \mathcal{M}_1^{(x_1)},
\end{equation}
where the operation $\mathcal{M}_i^{(x_i)}$ acts on $S_{i}$.
The  quantum operations of (selective) projective quantum measurements act as
\begin{equation}\label{SOPF}
    \mathcal{M}_j^{(x_j)} |\rho^{S})= |\Pi_{x_j}^j \rho^{S}  \Pi_{x_j}^j)=|\Pi_{x_j}^j)(\Pi_{x_j}^j| \rho^{S}  ),
\end{equation}
so the $|\Pi_{x_j}^j)$ in \eqref{SOPF} projects the maximally entangled states in \eqref{11} to the bipartite classical states. In this case, the $n$-point operation \eqref{11} can be briefly denoted as 
\begin{equation}\label{c13}
   (\Pi^{S_n}_{x_n}\otimes[\Pi^{S_{n-1}}_{x_{n-1}}\otimes\Pi^{A^{(n-1)}}_{x_{n-1}}]\otimes...\otimes[\Pi^{S_1}_{x_1}\otimes\Pi^{A^{(1)}}_{x_1}]|.
\end{equation}
We consider only  selective projective measurements throughout the paper. With these notations in hand, the joint  probability for $n$-point measurements can be expressed as 
\begin{equation}\label{NPMG}
    \mathcal{P}_{n:1}(x_{n:1}|\mathcal{M}_{n:1})= (\mathcal{O}_{n:1}^{(x_n:x_1)}|\mathcal{S}_{n:1}).
\end{equation}

Process states and process tensors can be established as equivalent representations. The precise mathematical correspondence between these two frameworks, particularly through the Choi-Jamiołkowski isomorphism within the operator-state formalism, is systematically presented in \cref{AnnA}. This unified perspective enables the treatment of multitime quantum processes as multipartite quantum states, thereby providing an effective framework for analyzing temporal correlations in LG inequalities.

A fundamental connection emerges through the realization that any process state can be expressed as the action of a process tensor on a sequence of maximally entangled states. This relationship is explicitly captured by the following operator-state correspondence:
\begin{equation}\label{T6}
    |\mathcal{S}_{n:1})=|\mathcal{T}_{n:1}[\bigotimes_{j=2}^{n}  \Phi^{A^{(j-1)}S_{j}}]),
\end{equation}
as detailed in \eqref{A6}. This formulation demonstrates that the process state essentially constitutes the Choi-state representation of the corresponding process tensor. The isomorphism preserves the complete information about temporal correlations while providing a state-like representation of quantum dynamical processes, enabling a unified treatment of spatial and temporal correlations.

\subsection{LG inequalities}
Having established the process-state framework, we now bridge it to the operational derivation of LG inequalities. The key lies in constructing joint probabilities \(P(Q_j, t_j; Q_i, t_i)\) from the process state $|\mathcal{S}_{n:1})$, which encode temporal correlations between measurements.

The general statement of macrorealism can be made operational by the following postulates \cite{LG,ELN14}:
\begin{enumerate}
 \item {\it Macrorealism per se}: A macroscopic object should have two or more macroscopically distinct states and at any time the object is in one of these states. 
 \item {\it Non-invasive measurability}: It is possible to measure the state without disturbing the subsequent dynamics. 
 \item {\it Induction}: The present state and the measurement outcome of the present state cannot be affected by future measurements.
\end{enumerate}
Based on these postulates, we can derive the LG inequalities for the two-point correlations of two-time measurements. Notice that in deriving LG inequalities, one only asks noninvasiveness of the performed measurements: If they hold true, then not performing a measurement  cannot be distinguished from averaging over their probabilities \cite{Plenio20}. In this case, there is no limit to the Markovianity of the process.

Let us consider a two-time measurement, or two measurements at times $t_i$ and $t_j$, with outcomes $Q_i$ and $Q_j$ respectively. The  joint pair probability for this two-time measurement to happen is $P(Q_j,t_j;Q_i,t_i)$. Then the corresponding two-time correlation function is
\begin{equation}
C(t_j,t_i)=\sum_{Q_j,Q_i}Q_jQ_iP(Q_j,t_j;Q_i,t_i).
\end{equation}
This $C(t_j,t_i)$ is a correlation function in the sense of classical probability theory. In quantum theory, we should consider instead the quantum measurements on the quantum system represented by the measurement operators $\hat{\mathcal{M}}_i$. Let $\rho$ be the density matrix of the quantum system which has the Markovian open quantum dynamics dictated by a Lindbladian $\mathcal{L}$. That is, we have the time evolution equation $\frac{d\rho}{dt}=\mathcal{L}\rho$ with a formal solution $\rho(t_j)=e^{\mathcal{L}(t_j-t_i)}\rho(t_i)$. If the quantum measurements are projective, i.e., $\hat{\mathcal{M}}(\cdot) =\sum_ma_m\Pi_m(\cdot) \Pi_m$, where $a_m=\pm1$ for simplicity and $\Pi_m$ are projection operators, then the quantum correlation function for the two-time measurement is
\begin{align}
C_q(t_j,t_i)=&\sum_{m,n}a_ma_n\text{tr}\bigl[\Pi_me^{\mathcal{L}(t_j-t_i)}(\Pi_n\rho(t_i)\Pi_n)\bigr]=\nonumber\\
=&\text{Tr}\bigl[\hat{\mathcal{M}}(t_j)\hat{\mathcal{M}}(t_i)\rho(t_i)\bigr],
\end{align}
where the $\hat{\mathcal{M}}(t_j)$ are written in the Heisenberg picture and in particular for $\hat{\mathcal{M}}(t_i)\rho(t_i)$  the evolution operator is simply an identity. In general, $C_q(t_j,t_i)$ is complex, and its imaginary part measures the noncommutativity of $\hat{\mathcal{M}}(t_j)$ and $\hat{\mathcal{M}}(t_i)$. By L\"uders' theorem \cite{Lud50}, we know that the non-invasiveness of measurements is equivalent to the commutativity of the measurement operators, so we should consider only the real part 
\begin{equation}
  C_{ji}=  \frac{1}{2}\Tr(\{\hat{\mathcal{M}}(t_j),\hat{\mathcal{M}}(t_i)\}\rho).
\end{equation}

To obtain the LG inequalities in the simplest form, we consider only the two-time measurements,  that is, $n=2$ in \eqref{10} and \eqref{11}. 
So  the two-time joint probability is
\begin{equation}\label{17}
    \mathcal{P}_{i,j}(x_i,x_j|\mathcal{M}_{i,j})= (\mathcal{O}_{i,j}^{(x_i,x_j)}|\mathcal{S}_{n:1}),\quad i\leqslant j.
\end{equation}
Since the quantum measurements affect both the quantum coherence of the system's state and its subsequent dynamics, the classical Kolmogorov consistency condition could be violated, to wit,
\begin{equation}
    \mathcal{P}_{i,j}(x_i,x_j|\mathcal{M}_{i,j})\neq \sum_{x_1,\ldots,\hat{ x_i},\ldots, \hat{x_j},\ldots, x_n}  \mathcal{P}_{n:1}(x_{n:1}|\mathcal{M}_{n:1})
\end{equation}
where the $\hat{x_i}$ are not summed over. In contrast, if the measurements do not affect the state of the system (i.e., macrorealism per se) and do not disturb the following dynamics of the system (i.e., noninvasive measurability), then we can obtain the two-time probabilities from the $n$-time joint probabilities and also the LG inequalities for the two-time probabilities. Notice that the induction condition is naturally satisfied in the present framework,
\begin{equation}
    \mathcal{P}_{i,j}(x_i,x_j|\mathcal{M}_{i,j})= \sum_{x_{j+1},\ldots, x_n}  \mathcal{P}_{n:1}(x_i,x_{n:j}|\mathcal{M}_{i}\otimes\mathcal{M}_{n:j})
\end{equation}
as the order of quantum operations matters.

In terms of the two-time probabilities \eqref{11}, the two-time correlation functions are
\begin{equation}
    C_{ij}=\sum_{x_i,x_j} x_i x_j \mathcal{P}_{i,j}(x_i,x_j|\mathcal{M}_{i,j}).
\end{equation}
Taking $n=3$ for example, if we can obtain all the two-time probabilities from the three-time joint probability. It is then obvious that 
\begin{align}\label{K3}
    K_3\equiv&C_{12}+C_{23}-C_{13}= \notag \\
    =&1-\sum_{x_i,x_j,x_k}(x_j-x_k)(x_j -x_i  )\mathcal{P}_{3:1}(x_i,x_j,x_k).
\end{align}
When $x_i,x_j,x_k=\pm 1$, we have 
\begin{equation}\label{25}
K_3\leqslant 1
\end{equation}
which represents one form of the standard Leggett-Garg (LG) inequalities. A stronger LG inequality, such as the Clauser-Horne form \cite{CH74,KP21}, can be expressed as
\begin{equation}\label{CHF}
    \text{CH}_3\equiv K_3+\abs{M_2- M_{2}^{(1)}}+\abs{M_3- M_{3}^{(1)}}+\abs{M_3- M_{3}^{(2)}},
\end{equation}
where the single-measurement statistics for the observable are defined by
\begin{align}
    M_{j}^{(i)}&=\sum_{x_i,x_j}  x_j \mathcal{P}_{i,j}(x_i,x_j|\mathcal{M}_{i,j}),\\
    M_{j}&=\sum_{x_j}  x_j \mathcal{P}_{j}(x_j|\mathcal{M}_{j}).
\end{align}

\section{Conditions for LG inequalities}\label{S3}

As mentioned above, the violation of LG inequalities could be influenced by different types of  dynamics and the decoherence of the open system. In order to see these influences clearly, we consider here the conditions on the process states for the LG inequalities to hold.

\begin{figure}
    \centering
   \includegraphics[width=0.48\textwidth]{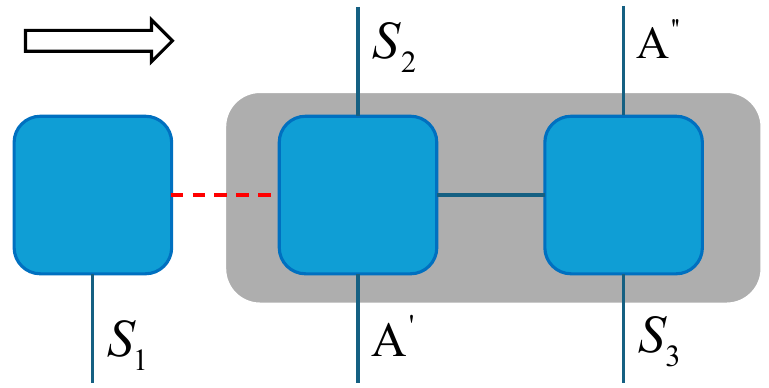}
    \caption{  Schematic of the process state \(|S_{3:1})\) as a tensor network. Arrows indicate temporal order, with \(S_1, S_2, S_3\) denoting system states at \(t_1, t_2, t_3\), and \(A', A''\) as auxiliary spaces. Gray shading highlights correlation between time steps. The red dashed line signifies retained system-environment correlations in the process state when initial system-environment correlations are present.}  
    \label{FIG1}
\end{figure}

Consider again  the case of three-time measurements with joint probability
\begin{equation}\label{TTMWJP}
    \mathcal{P}_{3:1}= (\Pi_{x_3}^{S_3}\otimes[\Pi_{x_2}^{S_2}\otimes\Pi_{x_2}^{A''}]\otimes[\Pi_{x_1}^{S_1}\otimes\Pi_{x_1}^{A'}]|\mathcal{S})
\end{equation}
where $\mathcal{S}$'s indices can be recovered as $\mathcal{S}^{S_3S_2A''S_1A'}$. The corresponding tensor network diagram is illustrated in \cref{FIG1}. We also have the probabilities for  two-time measurements:
\begin{align}
    \mathcal{P}_{1,2}=&\sum _{x_3} (\Pi_{x_3}^{S_3}\otimes[\Pi_{x_2}^{S_2}\otimes\Pi_{x_2}^{A''}]\otimes[\Pi_{x_1}^{S_1}\otimes\Pi_{x_1}^{A'}]|\mathcal{S}),\notag \\
    \mathcal{P}_{2,3}=&\sum_{x_1,y_1}(\Pi_{x_3}^{S_3}\otimes[\Pi_{x_2}^{S_2}\otimes\Pi_{x_2}^{A''}]\otimes[\Pi_{x_1y_1}^{S_1}\otimes\Pi_{x_1y_1}^{A'}]|\mathcal{S}), \notag \\
    \mathcal{P}_{1,3}=&\sum_{x_2,y_2}(\Pi_{x_3}^{S_3}\otimes[\Pi_{x_2y_2}^{S_2}\otimes\Pi_{x_2y_2}^{A''}]\otimes[\Pi_{x_1}^{S_1}\otimes\Pi_{x_1}^{A'}]|\mathcal{S}),\label{26}
\end{align}
where the double sum $\sum_{x,y}$ means that no projective measurement  has been done. Compared to \eqref{c13}, in the absence of a measurement the corresponding state remains to be a maximally entangled state
\begin{equation}
 (\sum_{x,y}\Pi_{xy}^{S}\otimes\Pi_{xy}^{A'}|= (\Phi^{A'S}|.
\end{equation}

Now suppose the  two-time probabilities can be obtained from the three-time joint probability \eqref{TTMWJP} as marginals, then we see that the conditions  in \eqref{26} are equivalent to satisfying one of 
\begin{align}
    (\Pi_{x_1y_1}^{S_1}|\mathcal{S}^{S_3S_2A''S_1A'})=&\delta _{x_1y_1}(\Pi_{x_1}^{S_1}|\mathcal{S}^{S_3S_2A''S_1A'}) \label{1A},\\
    (\Pi_{x_1y_1}^{A'}|\mathcal{S}^{S_3S_2A''S_1A'})=&\delta _{x_1y_1}(\Pi_{x_1}^{A'}|\mathcal{S}^{S_3S_2A''S_1A'}) \label{1B},
\end{align}
and one of 
\begin{align}
    (\Pi_{x_2y_2}^{S_2}|\mathcal{S}^{S_3S_2A''}_{x_1})=&\delta _{x_2y_2}(\Pi_{x_2}^{S_2}|\mathcal{S}^{S_3S_2A''}_{x_1}) \label{2A},\\
    (\Pi_{x_2y_2}^{A''}|\mathcal{S}^{S_3S_2A''}_{x_1})=&\delta _{x_2y_2}(\Pi_{x_2}^{A''}|\mathcal{S}^{S_3S_2A''}_{x_1}) \label{2B},
\end{align}
where $|\mathcal{S}^{S_3S_2A''}_{x_1})=(\Pi_{x_1}^{S_1}\otimes\Pi_{x_1}^{A'}|\mathcal{S})$ is the reduced process state after the first measurement.
We only need one condition from two conditions because in the operation part of the states, e.g., $\Phi^{S_1A'}$, are maximally entangled states.

The derived sufficient conditions (\ref{1A})-(\ref{2B}) possess a transparent physical interpretation: they rigorously enforce state classicality throughout the process dynamics. Specifically, condition (\ref{1A}) mandates that the initial state \(\rho^{S_1}_0\) must exhibit no quantum coherence in the measurement basis (i.e., remain strictly diagonal), while conditions (\ref{2A}) constrain the intermediate evolved state to maintain classical.  The conditions \eqref{1B} and \eqref{2B} imply that the coherence of input states does not affect subsequent outputs or dynamics. Violations of LG inequalities thus directly witness the failure of such classical constraints, linking non-macrorealism to quantum coherence in the temporal domain. This aligns with prior results on Markovian processes \cite{Plenio19}, but extends to non-Markovian scenarios through the process-state formalism.

The conditions \eqref{1A} restricts the initial state $\rho^{S_1}$ to be a classical state with diagonal density matrix. That is,  the process state should take the form of a ``quantum-classical'' state
\begin{equation}
|\mathcal{S}^{S_3S_2A''S_1A'})=\sum_{x_1}|\mathcal{S}^{S_3S_2A'A''}_{x_1})\otimes|\Pi^{S_1}_{x_1}).
\end{equation}
 If the conditions \eqref{1B}, \eqref{2A} and \eqref{2B} are furthermore satisfied, we have
\begin{equation}\label{333}
|\mathcal{S}^{S_3S_2A''S_1A'})=\sum_{x_1,x_2}|\mathcal{S}^{S_3}_{x_1,x_2})\otimes|\Pi^{S_2}_{x_2}\otimes\Pi^{A''}_{x_2})\otimes|\Pi^{S_1}_{x_1}\otimes\Pi^{A'}_{x_1}).
\end{equation}
Recalling the definition of process states (or the  diagrammatic rule \eqref{c8}), we know that from $A'$ to $S_2$ the system undergoes a time evolution $\mathcal{U}_1$, so that the particular form of \eqref{333} means this time evolution $\mathcal{U}_1$ should be a {\it classical} time evolution without generating quantum coherence on the system.
This is a very strong requirement for the time evolutions.
 
On the other hand, when only \eqref{1B} and \eqref{2A} are satisfied, the allowable process state takes the form\begin{widetext}
\begin{equation}\label{33}
|\mathcal{S}^{S_3S_2A''S_1A'})=\sum_{x_1,x_2}|\mathcal{S}^{S_3A''S_1}_{x_1,x_2})\otimes|\Pi^{S_2}_{x_2})\otimes|\Pi^{A'}_{x_1})
+\sum_{x_1}\left(|\mathcal{R}^{S_3S_2A''S_1}_{x_1})-|\mathcal{R}^{S_3S_2A''}_{x_1})\otimes|\Pi^{S_1}_{x_1}) \right)\otimes|\Pi^{A'}_{x_1}),
\end{equation}
where  the constituent elements satisfy the consistency conditions: $(\Pi_{x_1}^{S_1}|\mathcal{S}^{S_3A''S_1}_{x_1,x_2})=(\Pi^{S_2}_{x_2}|\mathcal{S}^{S_3S_2A''}_{x_1})$ and  $|\mathcal{R}^{S_3S_2A''}_{x_1})=(\Pi^{S_1}_{x_1}|\mathcal{R}^{S_3S_2A''S_1}_{x_1})$. Crucially, the components  $\mathcal{R}^{S_3S_2A''S_1}_{x_1}$ represent independent, freely adjustable states that define a general class of process-state solutions. 
The first summation term of \eqref{33} corresponds to a stronger condition $ (\Pi_{x_2y_2}^{S_2}|\mathcal{S}^{S_3S_2A''S_1A'})=\delta _{x_2y_2}(\Pi_{x_2}^{S_2}|\mathcal{S}^{S_3S_2A''S_1A'}) $ which is independent on the first measurement. It's physically requiring the first evolution segment to implement classical dynamics (e.g., dephasing channels). 
The second summation term represents the discrepancy between scenarios with and without the first measurement on  $S_1$, vanishing under the trace operation $(\Pi_{x_1}^{S_1}\otimes\Pi_{x_1}^{A'}|\mathcal{S})$. Physically speaking, this is equivalent to allowing the output of the first evolution to be non-classical and cause non-classical correlations in the process states, but such correlations must be canceled out by the first measurement.
If \eqref{1A} and \eqref{2A} are satisfied, we have 
\begin{equation}\label{34}
|\mathcal{S}^{S_3S_2A''S_1A'})=\sum_{x_1,x_2}|\mathcal{S}^{S_3A''A'}_{x_1,x_2})\otimes|\Pi^{S_2}_{x_2})\otimes|\Pi^{S_1}_{x_1})
 +\sum_{x_1}\left(|\mathcal{R}^{S_3S_2A''A'}_{x_1})-|\mathcal{R}^{S_3S_2A''}_{x_1})\otimes|\Pi^{A'}_{x_1}) \right)\otimes|\Pi^{S_1}_{x_1}),
\end{equation}
the first term of which requires that the initial state and the state output by the first evolution are both classical states. The second summation term relaxes the restrictions on the output of first evolution, allowing for non-classical correlations caused by the first evolution output, but such correlations must be canceled out by the first measurement. 

Likewise, we have other two possible process states: If \eqref{1B} and \eqref{2B} are satisfied, we have 
\begin{equation}\label{35}
|\mathcal{S}^{S_3S_2A''S_1A'})=\sum_{x_1,x_2}|\mathcal{S}^{S_3S_2S_1}_{x_1,x_2})\otimes|\Pi^{A''}_{x_2})\otimes|\Pi^{A'}_{x_1})
+\sum_{x_1}\left(|\mathcal{R}^{S_3S_2A''S_1}_{x_1})-|\mathcal{R}^{S_3S_2A''}_{x_1})\otimes|\Pi^{S_1}_{x_1}) \right)\otimes|\Pi^{A'}_{x_1}),
\end{equation}
the first term of which requires decoherence in both evolution segments, preventing initial coherence from propagating through subsequent dynamics. The second summation term relaxes the restrictions on the second evolution, allowing for the coherence of input states affect subsequent outputs or dynamics. But such effect also needs to be canceled out by the first measurement. Finally, when \eqref{1A} and \eqref{2B} hold:
\begin{equation}\label{3737}
|\mathcal{S}^{S_3S_2A''S_1A'})=\sum_{x_1,x_2}|\mathcal{S}^{S_3S_2A'}_{x_1,x_2})\otimes|\Pi^{A''}_{x_2})\otimes|\Pi^{S_1}_{x_1})
 +\sum_{x_1}\left(|\mathcal{R}^{S_3S_2A''A'}_{x_1})-|\mathcal{R}^{S_3S_2A''}_{x_1})\otimes|\Pi^{A'}_{x_1}) \right)\otimes|\Pi^{S_1}_{x_1}),
\end{equation}
demanding classical initial states combined with constrained second evolution that preventing coherence from propagating. And the second summation term, while relaxing constraints on the subsequent evolution, must be exactly canceled by contributions from the first measurement to maintain consistency.
These four distinct combinations of conditions  \eqref{1A}--\eqref{2B}  thus generate the family of "quantum-classical" process states \eqref{33}-\eqref{3737}, collectively denoted as $|\mathcal{S}_{QC})$. 
 
 As long as the process state satisfies one of these $|\mathcal{S}_{QC})$, the two-time probabilities can be obtained from a three-time probability as marginals, and one can derive the LG inequalities. If the process state is not any of the four types, then by the induction condition, we have
\begin{align}
    K_3=&1-\sum_{x_i,x_j,x_k}(x_j-x_k)(x_j -x_i  )\mathcal{P}_{3:1}(x_i,x_j,x_k)+\sum_{x_1,y_1,x_2,x_3} x_2 x_3 (\Pi_{x_3}^{S_3}\otimes[\Pi_{x_2}^{S_2}\otimes\Pi_{x_2}^{A''}]\otimes[\Pi_{x_1y_1}^{S_1}\otimes\Pi_{x_1y_1}^{A'}]|\Delta\mathcal{S})-\notag \\
    &-\sum_{x_1,x_2,y_2,x_3} x_1 x_3 (\Pi_{x_3}^{S_3}\otimes[\Pi_{x_2y_2}^{S_2}\otimes\Pi_{x_2y_2}^{A''}]\otimes[\Pi_{x_1}^{S_1}\otimes\Pi_{x_1}^{A'}]|\Delta\mathcal{S})\label{38}
\end{align}
where  $\Delta\mathcal{S}=\mathcal{S}-\mathcal{S}_{QC}$ is the deviation of the process state from the ``quantum-classical'' process state.

As for the stronger Clauser-Horne form, we must account for single-time measurement probabilities. Following the induction condition and \cref{CHF}, we have
\begin{align}
    \text{CH}_3=K_3+\left|\sum_{x_2} x_2 (I^{S_3}\otimes[\Pi_{x_2}^{S_2}\otimes\Pi_{x_2}^{A''}]\otimes\Phi^{A'S_1}|\Delta\mathcal{S}_1)\right|\notag\\
    +\left|\sum_{x_3} x_3 (\Pi_{x_3}^{S_3}\otimes\Phi^{A''S_2}\otimes\Phi^{A'S_1}|\Delta\mathcal{S}_1)\right|+\left|\sum_{x_3} x_3 (\Pi_{x_3}^{S_3}\otimes\Phi^{A''S_2}\otimes\Phi^{A'S_1}|\Delta\mathcal{S}_2)\right|,
\end{align}
\end{widetext}
where the state deviation  $|\Delta\mathcal{S}_i)=|\mathcal{S})-\mathcal{M}_i^{S_i}|\mathcal{S})$ quantifies the departure from the ``quantum-classical'' process state $\mathcal{M}_i^{S_i}|\mathcal{S})$, induced by the disturbance of measurement operation $\mathcal{M}_i^{S_i}=\sum_{x_i}|\Pi_{x_i}^{S_i})(\Pi_{x_i}^{S_i}|$.


These conditions \eqref{1A}--\eqref{2B}  are about the ability of obtaining the marginals from the joint probability. Our conditions generalize the no-coherence-generation requirement for LG inequalities in Markovian dynamics \cite{LLCC12}, and subsume the `classical process' criterion of \cite{KLCE20,KKCQ22} as a special case when $\mathcal{U}_1$ is a dephasing channel. 
Moreover, the ``quantum-classical'' form indicates possible relations to the results from \cite{Plenio20}. By \cite{Plenio20}, a classical process can be modeled by an initially classical system state  and a set of   non-coherence-generating-and-detecting  maps consisting of dynamical maps and completely dephasing maps. Here, we also allow arbitrary initial system states.
Since we have only considered  the projective measurements on the system, these conditions are stricter than the dephasing maps in \cite{Plenio20}. For non-Markovian processes, there exit memory effects in general, and hence the non-invasive measurability can be no longer true and the LG inequalities \eqref{25} could be violated. However, we can still learn these influences by comparing the particular process states/tensors with the above ``quantum-classical'' form.

\section{Violation of LG inequalities via deviation of process state}\label{S4}
In this section, we consider different influences on the violation of LG inequalities  by looking at the deviations from the ``quantum-classical'' conditions on process states introduced above. Our purpose here is to see if we could learn these influences by simply checking the form of process states/tensors.
 In the following subsections, when saying ``violating the LG inequalities'' we should keep in mind the \eqref{38} as a mean to detect various influences.

\subsection{State-disturbance influence on the violation of LG inequalities }\label{LGIMarkov}
We first show that the violation of  LG inequalities is clearly influenced by the disturbance  on the system's state caused by measurements. This problem has been studied thoroughly in \cite{Plenio19} where a one-to-one connection between satisfying LGI and Markovian dynamics that do not generate and detect coherence is established. Here, we revisit this result in the operator-state formalism.

Let us consider general quantum processes generated by a series of quantum operations $\mathcal{N}_i$, so that the corresponding open system evolution is $|\rho_j^S)=\mathcal{N}_{j}\circ \ldots \circ \mathcal{N}_{1}|\rho_0^S) $. In the absence of memory effects, the overall quantum process is a Markovian process. Now recall the Markovian condition for process tensors  \cite{MC}: A process tensor $ \mathcal{T}_{n:1}$ is Markovian if it can be written as 
\begin{equation}
    \mathcal{T}_{n:1}= \mathcal{T}_{n:n-1}\otimes \ldots \otimes \mathcal{T}_{2:1},
\end{equation}
which is equivalent to
\begin{equation}\label{PTMK}
    |\mathcal{S}_{n:1})= |\mathcal{S}_{n:n-1}^{A^{(n-1)}S_n}\otimes \ldots \otimes \mathcal{S}_{2:1}^{A'S_2}\otimes \rho_0^{S_1})
\end{equation}
where $\mathcal{S}_{i:i-1}^{A_iS_i}=\mathcal{N}_{i-1}^{S_i} (\Phi^{A^{(i-1)}S_i})$. As the quantum operation at each step is fixed, the particular form \eqref{PTMK} restricts the quantum process to follow a fixed form of dynamics. In other words, the influence here comes from the disturbance on the system state. Comparing \eqref{34} and \cref{PTMK}, we see that in both cases there is no entanglement between $S_1$ and $S_2$ and there is no entanglement between $S_2$ and $S_3$ which is also  independent on the first measurement; but unlike \eqref{34}, in \eqref{PTMK} there is no specification on projective measurements. Therefore, we can  study the influence from the disturbance on states in \eqref{PTMK}  by  checking the deviation from \eqref{34}.

By comparing \cref{PTMK} and \eqref{34}, we see that, for these two states to coincide upon the first two measurements, we require
\begin{align}
\rho_0^{S_1}=&\sum_{x_1} P_1(x_1)\Pi_{x_1}^{S_1},\label{39}\\
(\Pi_{x_1}^{A'}|\mathcal{N}_{1} |\Phi^{A'S_2})=&\sum_{x_2} P_2^{(x_1)}(x_2)\Pi_{x_2}^{S_2},\label{40}
\end{align}
where $P_1(x_1)=(\Pi_{x_1}^{S_1}|\rho_0^{S_1})$ and $P_2^{(x_1)}(x_2)=(\Pi_{x_1}^{A'}\otimes \Pi_{x_2}^{S_2}|\mathcal{N}_{1} |\Phi^{A'S_2})$. The condition \eqref{39} means that the input state for the  first measurement $\mathcal{M}_1$ is a diagonal classical state. Similarly, the condition \eqref{40} requires  that the input state for the  second measurement $\mathcal{M}_2$ is a diagonal classical state, after the first trace $(\Pi_{x_1}^{S_1}\otimes\Pi_{x_1}^{A'}|\mathcal{S})$  is taken. 
The condition \eqref{40} involves the traces because of the conditions on the second measurement do so.

If  the conditions \eqref{39} and \eqref{40} are satisfied, then the LG inequalities hold.  In general,  a quantum operation  $\mathcal{N}_1$  can generate coherence in such a way that the output state of $\mathcal{N}_1$ is not a diagonal classical state, even if the trace $(\Pi_{x_1}^{S_1}\otimes\Pi_{x_1}^{A'}|\mathcal{S})$ is taken.
Thus, we conclude that there is a positive  influence on the violation of LG inequalities for an  open system  from
 the state changes (e.g., the coherence generation) caused by the quantum operations.

\subsection{Non-Markovian influence on the violation of LG inequalities }
Now let us turn to the influence on the violation of LG inequalities from the memory effects. In the presence of memory effects, the process state cannot be written in the form of \eqref{PTMK}. Generally speaking, $|\mathcal{S}^{A^{(j)}S_{j+1}})$ and $|\mathcal{S}^{A^{(j-1)}S_{j}})$ are quantum correlated, thereby making it difficult for the total process state to conform to the ``quantum-classical'' form. Therefore, the memory effects have positive influences on the violation of LG inequalities, which is consistent with the results of previous work \cite{CA14}.

Notice that  there is no one-to-one connection between  memory effects and violation of LG inequalities, although  the memory effect will violate the postulate of non-invasive measurabilty. In fact, the failure of non-invasiveness could be witness by other more suitable ways such as violating the diagonal non-invasiveness \cite{Bud23}. By comparing the process states/tensors with the ``quantum-classical'' \eqref{33}--\eqref{3737} we can  just see the influence on the violation of LG inequalities, {\it instead of} witnessing memory effect by LG inequalities.

An interesting intermediate case between the memoryless and mnemonic processes is the process with finite Markov order, i.e. memories of finite length. The quantum Markov order has been characterized in the process tensor framework \cite{MO,MO2}, where a quantum process with finite-length Markov order has the process tensor of the following form
\begin{equation}
    \mathcal{T}_{HMF}=\sum_x P(x)\mathcal{T}_{H}^{(x)} \otimes \Delta_{M}^{(x)}  \otimes \mathcal{T}_{F}^{(x)} 
\end{equation}
where the subscripts $H,M,F$ label respectively the history, Markov order, and future parts of the quantum process.  The $\Delta_i^{(x)}$ is a particular quantum instrument such that $(O_M^{(x)}|\Delta_M^{(y)})=\delta_{xy}$. Since projective measurements are self-dual, we have $\Delta_{M}^{(x)}=\Pi_{M}^{(x)}$.
This form of process tensors allows a non-vanishing conditional mutual information $I(F:H|M)>0$, thereby indicating  memory effect. 
Using again \eqref{T6}, we obtain equivalently
\begin{equation}
    |\mathcal{S}_{HMF})=\sum_x P(x)|\mathcal{S}_{H}^{(x)} \otimes \mathcal{S}_M^{(x)}  \otimes \mathcal{S}_{F}^{(x)} ),
\end{equation}
where $|\mathcal{S}_M^{(x)})$ is Choi states of $ \Delta_{M}^{(x)} $.

To see the influence from  the Markov order part $M$, let us suppose the measurements do not affect the state of the system in the $H,M,F$ parts. In the three-time measurement case with now the Markov order-1 (i.e. the memory extends to the second-step time evolution), we have the following property  of the marginal probabilities (cf. Eq. (4) of \cite{MO})
\begin{equation}\label{MOProb}
    \frac{\mathcal{P}_{2,3}(x_2,x_3|\mathcal{M}_{2,3})}{\mathcal{P}_{2}(x_2|\mathcal{M}_{2})}=\frac{\mathcal{P}_{3:1}(x_{3:1}|\mathcal{M}_{3:1})}{ \mathcal{P}_{1,2}(x_1,x_2|\mathcal{M}_{1,2})},
\end{equation}
where $\mathcal{P}_{2}(x_2|\mathcal{M}_{2})=\sum_{x_3}\mathcal{P}_{2,3}(x_2,x_3|\mathcal{M}_{2,3})$ according to the induction condition.
Since the instrument in the Markov order part $M$ is chosen such that $(\mathcal{O}^M_{y}| \Delta_M^{(y)})=\delta_{xy}$, the second measurement in $M$  effectively acts like an  identity matrix (cf. the double summation in \eqref{26}), that is
\begin{equation}
   \sum_{x} |\Pi^M_{x})(\Pi^M_{x}|\mathcal{S}_{3:1})=\sum_{x,y}|\Pi^M_{xy}) (\Pi^M_{xy}|\mathcal{S}_{3:1})={\bf1}^M|\mathcal{S}_{3:1}).
\end{equation}
Therefore, $\mathcal{P}_{1,3}$ can still be derived from $ \mathcal{P}_{3:1}$ as marginal
\begin{equation}
    \mathcal{P}_{1,3}(x_1,x_3|\mathcal{M}_{1,3})= \sum_{x_2}\mathcal{P}_{3:1}(x_{3:1}|\mathcal{M}_{3:1}).
\end{equation}
$\mathcal{P}_{2,3}$, on the other hand, can be obtained from $\mathcal{P}_{1,3}$ and $\mathcal{P}_{1,2}$ with the help of  \eqref{MOProb}. Using these probabilities, we can obtain\begin{widetext}
\begin{align}
    K_3=&\sum_{x_i,x_j,x_k}\Bigl[x_ix_j \mathcal{P}_{3:1}(x_i,x_j,x_k) +x_jx_k \frac{\mathcal{P}_{3:1}(x_i,x_j,x_k)}{\mathcal{P}_{1,2}(x_i,x_j)}\mathcal{P}_{2}(x_j)-x_ix_k \mathcal{P}_{3:1}(x_i,x_j,x_k)\Bigr ]\nonumber\\
    \doteq&1-\sum_{x_i,x_j,x_k}(x_j-x_k)(x_j \mathcal{P}_{2}(x_j)-x_i \mathcal{P}_{1,2}(x_i,x_j) ) \frac{\mathcal{P}_{3:1}(x_i,x_j,x_k)}{\mathcal{P}_{1,2}(x_i,x_j)}\label{K3orig}
\end{align}
\end{widetext}
where we have used \eqref{MOProb} in the first line and 
 in the second line (with $\doteq$) the relation
\[
    \sum_{x_j,x_k}x_j^2 \mathcal{P}_{2,3}(x_j,x_k)=\sum_{x_j,x_k} \mathcal{P}_{2,3}(x_j,x_k)=1
\]
for {\it dichotomic outcomes} $x_j=\pm1$ has been applied. When $\mathcal{P}_{2}(x_j)<  \mathcal{P}_{1,2}(x_i,x_j) $, the LG inequalities could be violated. In this case,  we already see that  the influence on LG inequalities from the memory effect is not in a one-to-one manner.

\subsection{An example}\label{S5}
To exemplify the above general considerations, we consider here a simple noisy model with two-qubits which has been previously considered in \cite{NBS20}. In this model, the initial state of the  system is $\rho_0$, and the initial state  vector of the  environment is also $\ket{\psi_E}=(\ket{+}+\ket{-})/\sqrt{2}$, where the $\ket{\pm}$ are just the qubit states. The total Hamiltonian for the $\rho^{SE}$ is 
\begin{equation}
    H=\omega (\ket{+-}\bra{-+}+\ket{-+}\bra{+-}).
\end{equation}

Let us consider two steps of time evolutions generated by $H$: The time interval for the first step is 
$[0, t_1=\tau_1]$, and the time interval for the second step is $[t_1, t_2=t_1+\tau_2]$. Suppose the measurements at $t_1$ and $t_2$ are projective measurements.
 Then the process state for this two-time measurement is
\begin{equation}\label{54}
    |\mathcal{S}^{S_3S_2A''S_1A'})=|\rho_0^{S_1}\otimes\frac{1}{2}[ \Pi^{S_3S_2A''A'}_{\psi_+}+ \Pi^{S_3S_2A''A'}_{\psi_-}] )
\end{equation}
where $\Pi^{S_3S_2A''A'}_{\psi_-}$ is the density matrix of the state 
\begin{widetext}
\begin{align}
    2\sqrt{2} \ket{\psi_-^{A'S_2A''S_3}}=&\ket{----}
   +\cos\theta_1 \ket{++--}+i\sin \theta_1 \ket{-+--}
   +\cos\theta_2 \ket{--++}+i\sin \theta_2 \ket{++-+}+\notag \\
  & +\cos\theta_1\cos\theta_2 \ket{++++}+i\cos\theta_1\sin \theta_2 \ket{---+}
   +i\sin\theta_1\cos\theta_2 \ket{-+++}-\sin\theta_1\sin \theta_2 \ket{+--+}
\end{align}
and  $\Pi^{S_3S_2A''A'}_{\psi_+}$ is the density matrix of 
\begin{align}
    2\sqrt{2} \ket{\psi_+^{A'S_2A''S_3}}=&\ket{++++}
   +\cos\theta_1 \ket{--++}+i\sin \theta_1 \ket{+-++}
   +\cos\theta_2 \ket{++--}+i\sin \theta_2 \ket{--+-}+\notag \\
   &+\cos\theta_1\cos\theta_2 \ket{----}+i\cos\theta_1\sin \theta_2 \ket{+++-}
   +i\sin\theta_1\cos\theta_2 \ket{+---}-\sin\theta_1\sin \theta_2 \ket{-++-}.
\end{align}
Here, $\theta_i=\omega \tau_i$ and we have decomposed the time evolutions $e^{-iHt}$ into the triangular functions.

Generally speaking, this process state \eqref{54} cannot be written in a Markovian form \eqref{PTMK}, except for the case with $\theta_1=k\pi$ in which we have
\begin{equation}
    2\sqrt{2} \ket{\psi_-^{A'S_2A''S_3}}=\bigl(\ket{--}+(-1)^k\ket{++}\bigr)
  \otimes\bigl(\ket{--}+\cos \theta_2\ket{++}+ i (-1)^k \sin \theta_2\ket{-+}\bigr ),
\end{equation}
and hence the corresponding process state
\begin{equation}\label{58}
    |\mathcal{S}^{S_3S_2A''S_1A'})=|\rho_0^{S_1}\otimes \Phi^{S_2A'}_k \otimes\frac{1}{2}[\Pi^{S_3A''}_{\psi_+}+ \Pi^{S_3A''}_{\psi_-}] ),
\end{equation}
with $\ket{\Phi^{S_2A'}_k}=(\ket{--}+(-1)^k\ket{++})/\sqrt{2}$, takes the form of \eqref{PTMK}. So we let 
$\rho_0^{S_1}=\sum_{x_1} P(x_1) \Pi_{x_1}^{S_1}$ in this case, then after the first step it becomes
$
    \rho_1^{S_2}=(\Phi^{S_1A'} |\mathcal{S}^{S_2S_1A'})=\sum_{x_1} P(x_1) \Pi_{x_1}^{S_2}
$
with $\Pi_{x_1}^{S_2}=( \Pi_{x_1}^{A'}|\Phi^{S_2A'}_k)$. Therefore, we can rewrite \eqref{58} as
\begin{equation}
    |\mathcal{S}^{S_3S_2A''S_1A'})=\sum_{x_1} P(x_1) P'(x_1) |\Pi_{x_1}^{S_1}\otimes  \Pi_{x_1}^{S_2}\otimes\Pi_{x_1}^{A'}\otimes \mathcal{S}^{S_3A''})
    +\sum_{x_1} P(x_1) |\Pi_{x_1}^{S_1}\otimes (\Phi^{S_2A'}_k- P'(x_1) \Pi_{x_1}^{S_2}\otimes\Pi_{x_1}^{A'})\otimes \mathcal{S}^{S_3A''}),
\end{equation}
where $P'(x_1) =(\Pi_{x_1}^{S_2}\otimes \Pi_{x_1}^{A'}|\Phi^{S_2A'}_k)$.  If we choose the measurement basis for two measurements as
$\{\ket{x_1}_{S_1}\}$ and  $\{\ket{x_1}_{S_2}\}$ respectively, the state has the ``quantum-classical'' form, thereby satisfying the LG inequalities and reflecting classical macrorealism.

When $\theta_1\neq k\pi$, for example, $\theta_1=\theta_2=(k+1/2)\pi$, we have
\begin{equation}\label{thkp1}
    2 \ket{\psi_-^{A'S_2A''S_3}}=i(-1)^{k-1}\ket{+-}_{S_2A''}\otimes \ket{ \Phi_0}_{A'S_3}
 +\ket{--}_{S_2A''}\otimes \ket{ \Phi_1}_{A'S_3}.
\end{equation}\end{widetext}
Then the corresponding process state is
\begin{equation}
    |\mathcal{S}^{S_3S_2A''S_1A'})=|\rho_0^{S_1}\otimes\frac{1}{2}[ \Pi^{A'S_2A''S_3}_{\psi_+}+ \Pi^{A'S_2A''S_3}_{\psi_-}]),
\end{equation}
from which we can easily obtain the reduced process state
\begin{equation}\label{EE1}
    |\mathcal{S}^{S_2S_1A'})=|\rho_0^{S_1}\otimes\frac{1}{2}[ \Pi^{A'S_2}_{\phi_+}+  \Pi^{A'S_2}_{\phi_+}])
\end{equation}
where
\begin{align}
    \ket{\phi_+}=&\ket{+}_{A'}\otimes(\ket{+}+i(-1)^{k-1}\ket{-})_{S_2},\notag \\
    \ket{\phi_-}=&\ket{-}_{A'}\otimes(\ket{-}+i(-1)^{k-1}\ket{+})_{S_2}.
\end{align}
As before, we assume that $\rho_0^{S_1}=\sum_{x_1} P(x_1) \Pi_{x_1}^{S_1}$, then  after the first step it becomes
\begin{equation}
    \rho_1^{S_2}=(\Phi^{S_1A'} |\mathcal{S}^{S_2S_1A'})=P'(+) \Pi_{\phi_+}^{S_2}+P'(-) \Pi_{\phi_-}^{S_2}.
\end{equation}
In order to discard the influence from the state disturbance, we consider the particular measurement bases
$\{\ket{x_1}_{S_1}\}$ and $\{\ket{\phi_+}_{S_2}, \ket{\phi_-}_{S_2}\}$ for two measurements. So we have 
\begin{equation}\label{65}
    (\Phi^{S_2A''} |\mathcal{S}^{S_3S_2A''S_1A'})=|\rho_0^{S_1}\otimes \Phi^{S_3A'}).
\end{equation}
We further let $\rho_0^{S_1}=\left(
    \begin{array}{cc}
     a & c \\
     c^* & b \\
    \end{array}
    \right)$ in the basis $\{\ket{-},\ket{+}\}$. Then \eqref{65} becomes
\begin{equation}
    (\Phi^{S_1A'} |\mathcal{S}^{S_3S_2A''S_1A'})=(\rho_+^{S_3S_2A''} +\rho_-^{S_3S_2A''}),
\end{equation}
where $\rho_{\pm}^{S_3S_2A''}=\left(
    \begin{array}{cc}
     a & c \\
     c^* & b \\
    \end{array}
    \right)$ in the respective bases $\{\ket{\phi_-}_{S_2}\otimes\ket{--}_{A''S_3},i(-1)^{k-1}\ket{\phi_+}_{S_2}\otimes\ket{-+}_{A''S_3}\}$ and $\{i(-1)^{k-1}\ket{\phi_-}_{S_2}\otimes\ket{+-}_{A''S_3},\ket{\phi_+}_{S_2}\otimes\ket{++}_{A''S_3}\}$.
    If the measurement bases for the first and third measurements are the same as $\{\ket{x_1}_{S_1}\}$ with
\begin{equation}
    \left(
\begin{array}{c}
 \ket{+}_{S_1} \\
 \ket{-}_{S_1} \\
\end{array}
\right)= \left(
\begin{array}{cc}
 \cos \theta & -i\sin \theta \\
 i\sin \theta & \cos \theta \\
\end{array}
\right)
\left(
\begin{array}{c}
    \ket{+} \\
    \ket{-}\\
\end{array}
\right),
\end{equation}
and the second measurement basis is $\{\ket{\phi_+}_{S_2}, \ket{\phi_-}_{S_2}\}$. Here $\theta\in[-\pi,\pi]$ is just a parameter.
We then compute the probabilities
\begin{align}
    &\mathcal{P}_{1,2}(+,\phi_+)=\mathcal{P}_{2,3}(\phi_+,+)= \mathcal{P}_{1}(+)\cos^2 \theta,\notag \\
    &\mathcal{P}_{1,2}(+,\phi_-)=\mathcal{P}_{2,3}(\phi_-,+)= \mathcal{P}_{1}(+)\sin^2 \theta, \notag \\
    &\mathcal{P}_{1,2}(-,\phi_-)=\mathcal{P}_{2,3}(\phi_-,-)= \mathcal{P}_{1}(-)\cos^2 \theta,\notag \\
    &\mathcal{P}_{1,2}(-,\phi_+)=\mathcal{P}_{2,3}(\phi_+,-)= \mathcal{P}_{1}(-)\sin^2 \theta, \notag \\
    &\mathcal{P}_{1,3}(x_1,x_1)= \mathcal{P}_{1}(x_1)\cos^2 2\theta,\notag \\
    &\mathcal{P}_{1,3}(x_1,-x_1)= \mathcal{P}_{1}(x_1)\sin^2 2\theta.
\end{align}
Thus we have 
\begin{equation}
    K_3=2\cos 2\theta -\cos 4\theta,
\end{equation}
which is $>1$ so long as $\cos2\theta>0$. Unlike \cref{58} that strictly satisfies quantum-classical conditions, the entanglement between subsystems \(S_2\) and \(A''\) established in \cref{thkp1} generates non-Markovian memory effects. This enables anomalous enhancement of temporal correlations (\(K_3 > 1\)), where the observed violation magnitude directly quantifies the non-classical temporal ordering emerging from coherent quantum evolution.

\section{Conclusion and discussion}\label{S6}
In this paper, we have studied the LG inequalities in the framework of the operator-state formalism. We have exploited the process tensor in its Choi-state form, which we also call the process state, to investigate the LG inequalities. In particular, we have derived a set of sufficient conditions on the process states for the LG inequalities to hold. We have exploited these conditions on the process states to investigate different influences on  the LG inequalities by comparing the structure of process states/tensors. 

The use of process states highlights the state structure in testing  macrorealism. In spite of the existence of various influences from environment on the LG inequalities for the open system, it is found that the violation of LG inequalities (or other inequalities alike) is still a witness for the existence of quantum coherence in the open system \cite{Plenio19,Plenio20,SD19}. In terms of the ``quantum-classical'' process states, we can clearly see whether there is quantum coherence in this temporal setting. As for the influence from memory effects, there is no one-to-one  relation between existence of memory effect and the violation of LG inequalities. Nevertheless, we can still see the influence by comparing the state structures, which is of practical importance in realistic setups.

We have mainly considered the sufficient conditions for the LG inequalities with three-time measurements. Such conditions can be readily generalized to more time measurements. In deriving these conditions we obtain the marginals of a temporal process state. It then seems possible to construct new inequalities testing the genuine temporal nonclassicality by using these marginals in  analogy to the multipartite Bell scenario (e.g. \cite{JDGSM18}). This analogy is of course incomplete as the temporal processes are time-ordered so that the partitions into time sub-intervals require new considerations.

The problem of the existence of a joint probability over time, when the marginals at each time are known, is not an easy one in quantum theory. There exists a no-go result \cite{nogo} stating that it is impossible to do so in quantum theory. Recently in \cite{FP22}, a new type of quantum states over time allowing the existence of joint probability over time is constructed by using the Jordan product of two Choi states in the forward and backward time directions. Here, instead of Jordan product, we use the Choi states in two time directions to rewrite the process tensors as a multiparite quantum state. This way, the temporally extended quantum states are rewritten as a zigzag-time-ordered quantum state. Whether the process states evade the no-go arguments deserves further consideration.

We finally remark that both the violation of  LG inequalities \cite{Exp1} and the process tensors \cite{Exp2} can be experimentally realized in the quantum optical platform. So, the process states considered in this work may also be experimentally realizable.
\begin{acknowledgments}
    Z.H. is supported by the National Natural Science Foundation of
 China under Grant No. 12305035. X.G. is supported by Yancheng Institute of Technology
(xjr2024030).
 X.G. also
 would like to thank James Fullwood for helpful discussions.
\end{acknowledgments}
\appendix
\section{Process tensors and process states}\label{AnnA}
  A process tensor \cite{PT,PTR2} is a multilinear map $\mathcal{T}_{n;0}:{\bf A}_{n-1;0}\mapsto\rho_n$ from a set ${\bf A}_{n-1;0}$ of $(n-1)$ quantum operations to the final state $\rho_k$ obtained after these $(n-1)$ operations.  Let us denote by $A_i\in{\bf A}_{n-1;0}$ the quantum operation at the $i$-th step/time and by $\mathcal{U}_{_{j:i}}$ the unitary evolution $U_{j:i}(\cdot)U_{j:i}^\dag$. Then the process tensor  $\mathcal{T}_{n;0}$ corresponds to the global evolution 
$\rho_n^{SE}=\mathcal{U}_{n,n-1}A_{n-1}\mathcal{U}_{n-1,n-2}...A_1\mathcal{U}_0A_0\rho_0^{SE}$, so
\begin{equation}\label{4}
\rho^S_n=\text{tr}_E(\rho_n^{SE})=\mathcal{T}_{n;0}[{\bf A}_{n-1;0}]=\sum_{{\bf s},{\bf x}}\mathcal{T}_{{\bf s},{\bf x}}{\bf A}_{{\bf s},{\bf x}},
\end{equation}
where
\[
\mathcal{T}_{{\bf s},{\bf x}}=\sum_{{\bf e}}\prod_{j=0}^{k-1}\Bigl(\mathcal{U}_{j+1;j}\rho_0^{SE}\Bigr)_{{\bf s},{\bf x}},\quad{\bf A}_{{\bf s},{\bf x}}=\prod_{j=0}^{k-1}\Bigl(A_{j}\Bigr)_{{\bf s},{\bf x}}
\]
with the collections ${\bf s,x,e}$ standing for the system, operation, environment (matrix) indices respectively. 
By  the Choi-Jamio\l kowski isomorphism, the quantum operations $A_j:\mathcal{H}_i\rightarrow\mathcal{H}_o$ can be transformed into the corresponding Choi states $O_j=A_j\otimes I(\Phi)$, where $\Phi$ is a projection to the maximally entangled states $\sum_i\ket{ii}$ in the Hilbert space $\mathcal{H}_o\otimes\mathcal{H}_i$. Denote by the nested sequence of Choi states by  $O_{k:0}^{x_{k:0}}$, then the joint probability of the whole quantum process producing outcomes $\{x_{j}\}\equiv x_{k:0}$ is
\begin{equation}\label{5}
P_{k:0}(x_{k:0}|{\bf A}_{k-1:0})=\text{tr}\Bigl(O_{k:0}^{x_{k:0}}\mathcal{T}_{k:0}\Bigr).
\end{equation}
This is a generalization of the familiar Born rule $p_i=\text{tr}(\Pi_i\rho)$ with projector $\Pi_i$.

The process tensor for $(n-1)$ steps of evolutions, as defined in \eqref{4}, can be expressed as an operator/tensor state
\begin{equation}\label{IPDPT}
    \mathcal{T}_{n:1}= (  I^E|\mathcal{U}_{n,n-1}\circ \ldots\circ \mathcal{U}_{2,1}|\rho^{SE}_0), 
\end{equation}
where $(I^E|$ means taking the trace over the environment $E$. The operator state for $(n-1)$-time measurements with outcomes $x_i,i=1,...,n-1$ is 
\begin{align}
  & |\mathcal{O}_{n:1}^{(x_n:x_1)})= O_{n,n-1}^{(x_{n-1})}\otimes \ldots \otimes O_{2,1}^{(x_1)}|I^S)=\nonumber\\
   =& (\mathcal{M}_{n-1}^{(x_{n-1})}\otimes I_{n-1})\otimes \ldots \otimes (\mathcal{M}_1^{(x_1)}\otimes I_1)|I^S )
\end{align}
where $O_j^{(x_j)}=\mathcal{M}_j^{(x_j)}\otimes I(\Phi)$.
Finally, the joint probability \eqref{5} can be expressed as 
\begin{equation}
    \mathcal{P}_{n:1}(x_{n:1}|\mathcal{M}_{n:1})= (\mathcal{O}_{n:1}^{(x_n:x_1)}|\mathcal{T}_{n:1}).
\end{equation}

Comparing the process tensors with the process states defined in the main text, we see that they are  in fact the same thing.    A process state can be  considered as a process tensor acting the tensor product of maximally entangled states,
\begin{equation}\label{A6}
    |\mathcal{S}_{n:1})=|\mathcal{T}_{n:1}[\bigotimes_{j=2}^{n}  \Phi^{A^{(j-1)}S_{j}}])
\end{equation}
if $S_j$ denotes the system before an operation $O$ and $A^{(j)}$ denotes the system after an operation $U$. Namely, we pull the curves in \eqref{c8} straight for \eqref{A6} to hold. This is nothing but the Choi-state form of the process tensor.
So the final $n$-point probabilities have the same structure.


\end{document}